# Switchable Single/Dual Edge Registers for Pipeline Architecture

Suyash Vardhan Singh and Rakeshkumar Mahto
Computer Engineering Program
California State University,
Fullerton, CA 92831

*Abstract*— **The demand for low power processing is increasing due to mobile and portable devices. In a processor unit, an adder is an important building block since it is used in Floating Point Units (FPU) and Arithmetic Logic Units (ALU). Also, pipeline techniques are used extensively to improve the throughput of the processing unit. To implement a pipeline requires adding a register at each sub-stage that result in increasing the latency. Moreover, designing a low power pipeline adder with low latency has drawn a lot of attention.**

**In a pipelined architecture that uses Dual Edge Triggered (DET) based registers can help in reducing the latency since they can capture input data at both clock edges. However, for high input activity, a DET flip-flop consumes more power than a Single-Edge Triggered (SET) flip-flop. Moreover, it is required to replace each Flip-Flop (FF) in the processor with Dual Edge Triggered (DET) Flip-Flop which will be a considerable area and power overhead. Therefore, it is desirable to have a switchable DET to SET depending on input activity or load condition to reduce the dynamic power consumption.**

**In this paper, we are proposing a new shift register which imitates DET FF based shift register without the need of special DET FF. The proposed shift register improved the latency in a 4-bit pipelined adder by two-fold. Additionally, the power delay product was reduced by 44.16%.**

I. INTRODUCTION

As technology is moving forward, feature size continues to shrink, and additional transistors are added following Moore's law. In order to continue scaling to reduce power consumption, the voltage also needs to be scaled by technology[1].  In recent times, the demand for low power computing in portable electronics, mobile devices, and wearable device etc. is on the rise.  For any processor, an adder is one of the main building blocks. It participates in arithmetic logic unit, floating point unit, cryptography and many more. Most of the processors which are in use today, extensively use pipeline technique to improve the throughput. The implementation of a low power adder with low latency in a pipeline architecture is still a challenging problem. Improving the performance of pipelined adder circuits will improve the overall system performance.

One of the most critical aspects of designing a low power processor is dynamic power. Various techniques are used to reduce power consumption. One such technique is clock gating [2],[3] where a portion of the D-flip flop is shut off thereby not switching states. This results in a reduction of dynamic power. Another approach of reducing dynamic power is by using pre-defined dual voltage source (Vdd) and dual threshold voltage (Vt) [4]. By using clock gating and dual threshold voltage (Vt) technique in a  4-bit BCD adder, average power in a 4-bit BCD adder was reduced by 62.8% [5]. Additionally, dynamic scaling of voltage and frequency [6] with changing load conditions are among a few other techniques used in portable embedded systems that can increase throughput and lower the dynamic power as per load requirement. Dynamic scaling of the frequency (DSF) is done through PLL which multiply the low frequency clock generated from the crystal oscillator. Dynamic scaling of the frequency and Dynamic Scaling Voltage (DSV) techniques were applied [7] on the sparse module $2^{n+1}$ Adder. Experimental data revealed that using DVS and DFS techniques lowered the power by 11.4% in sparse module $2^{n+1}$ Adder.

The throughput and latency are the other essential factors to consider while designing a low power processor. The throughput can be improved by utilizing a pipelined architecture. For implementing a pipeline, the task is divided into subtasks. Pipeline registers are added between each subtask. This addition of registers at each stage increases the latency since most of these registers are single edge triggered (SET) FF. Latency issue can be improved by using Dual Edged Triggered (DET) flip-flop (FF) based register. The DET flip-flop processes data at both rising and falling edge as compared to regular SET flip-flop. It is shown that DET FF is more energy efficient as compared to the SET FF [8]. Implementing a pipelined architecture on an adder circuit with DET FF based registers can mitigate pipeline latency and power consumption issues to some extent when compared to a conventional design.

Recently, many novel design methods have been proposed for DET FF to speed up the system with capturing data at both the clock edges. In work [9]–[11] DET FF is built using a lesser number of transistors to save power and area. In another technique [12]–[14], the authors generated a pulse at both edges for creating a DET FF. However, both the techniques have an area overhead. In the latter technique, an increase in switching activity increases dynamic power consumption. It is also presented in [15], a high input activity to the DET FF may lead to higher power dissipation than single edged one. Instead of just DET FF, it is desirable to build a technique that can dynamically switch between DET to Single Edge Triggered (SET) Flip Flop depending on workload. In this work, we present a technique that can be used in an existing pipelined adder architecture with minimum area overhead. The technique allows dynamically switching between SET FF and DET FF without the need of a special dual-edged FF or a pulse generating circuit. The proposed technique can be easily implemented for reducing the latency without the requirement of replacing all the D-Flip Flop with DET FF.

The organization of rest of the paper is as follows. Section II describes the proposed shift register circuit and its comparative simulation with a conventional shift register. In Section III the implementation of the shifter circuit on a 4-bit pipelined parallel adder circuit is presented. The conclusion of this paper is discussed in Section IV.

II. SINGLE/DUAL EDGED SHIFT REGISTER

Inter-modular communication design is part of the overall System on Chip (SOC) design. There are two techniques used for data transfer between two modules, parallel data transmission and bit serial data transmission. In parallel data transmission technique data is sent through a bus. Whereas in serial data transmission technique the data is sent one bit at a time through a single wire. Although the bit serial data transmission is better compared to the parallel data transmission in terms of area, less leakage, fewer driver and better routability [16], [17], they are bounded by the clock frequency. Since the data moves from input to output terminal synchronously at one of the clock edges only this reduces the throughput. To solve this problem, the DET flip-flop can be used since they transfer data from input to output at both edges thereby increasing the throughput for the same frequency. For implementing DET flip-flop different techniques have been used in past that can be classified in three different categories 1) Conventional dual edge triggered flip-flops, 2) explicit pulsed DET flip-flop and 3) implicit pulsed flip-flops [8]. The conventional DET flip-flop utilizes negative and positive edge flip-flop using a multiplexer that let these FFs operate at adjacent clock edges. This technique is non-ideal since it requires more area and increase in switching activity that results in higher power consumption. The explicit pulsed DET flip-flop uses the pulse generating circuit to generate a pulse at both the rising and falling edges. Nevertheless, this technique imposes higher power consumption. Whereas, the implicit pulsed FF utilizes clock and delayed clock into two series device embedded in the latching part. However, this technique has lower performance due to deeper N-Type Metal Oxide Semiconductor (NMOS) stack and power overhead due to the pulse generating circuit.

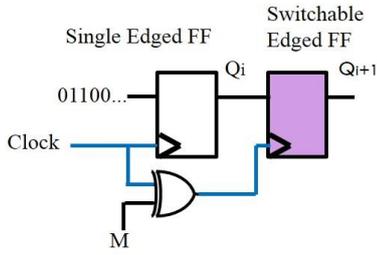

Figure 1 Switchable DET and SET FF register pair

However, none of the techniques in all the three categories can be switch between SET FF and DET FF based on load requirement and input switching activity to save power. Moreover, for reducing the latency of pipeline DET FF will be required to replace all the D-Flip Flops in the design which has huge area overhead. Whereas the techniques presented in this paper uses the existing D-Flip Flop based register to mimic DET FF based register.

### A. Proposed Switchable Register

The proposed technique uses a pair of D-flip flops as a register where the first one work as single edged (rising edge) flip-flop (SET FF) whereas the other acts as a switchable single edged flip-flop as shown in Figure 1. When the signal M is set to low, then both flip-flops work on the same edge (rising edge). Whereas, when the signal M is set to a high, this results in clock inversion to the second FF which makes it capture data at the falling edge of the original clock signal. In this condition the flip-flop pair transfer data from input to output at both edges.

When M='1', the first D-Flip flop holds the data until the falling edge is detected. As soon as falling edge is detected by the next switchable flip-flop, it reads the data and hold it until the arrival of next falling edge.

These pair of adjacent edge triggered flip-flop imitate a dual edge-triggered flip-flop. The basic operation and truth table is shown in Table 1.

**Table 1 Truth Table**

| M | Edge Switchable D-Flip Flop | Flip Flop pair |
|---|---|---|
| 0 | Rising Edge | Single edge shift register |
| 1 | Falling Edge | Dual edge shift register |

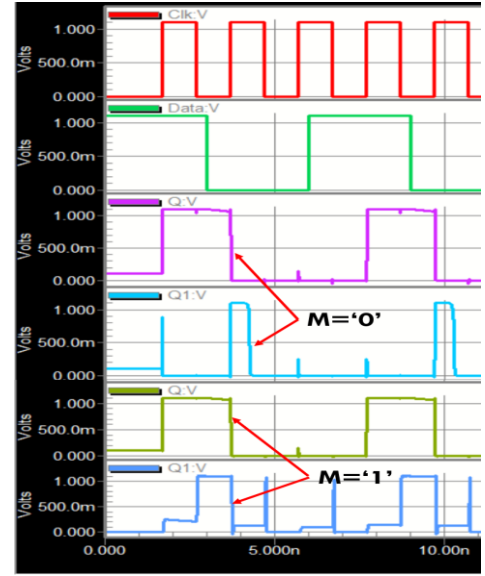

Figure 2 The SPICE simulation of shifter circuit

The minimum clock period for this configuration can be calculated by determining worst case logical delay between two adjacent registers. This is called critical delay. Let's assume the critical delay to be equal to $T_{\text{Max\_logic}}$. The setup and hold time for the registers are $T_{setup}$ and $T_{hold}$ respectively. The minimum clock period required for the sequential circuit is given by

$$\frac{T}{2} \geq T_{C2Q} + T_{Max\_logic} + T_{setup} - T_{Xor\_gate}$$

For satisfying the hold time condition the contamination delay ($T_{Min\_logic}$) or the minimum propagation delay through logic network plays a very important role. The hold time is given by,

$$T_{hold} \leq T_{C2Q} + T_{Min\_logic} - T_{Xor\_gate}$$

Where, $T_{Xor\_gate}$ is propagation delay across the XOR gate.

### B. SPICE Simulation Result

The SPICE simulation for both the cases, SET FF and DET are shown in Figure 2 is performed using the Berkeley Predictive Technology (BPTM) in a 45nm CMOS technology [18]. The Clock to Q (C2Q) delay from clock at the normal D Flip Flop to the output terminal of the switchable flip-flop is improved by 95.1%.

The average power consumption of the modified shifter was measured to be 40.3 µwatts for clock operating at 250 MHz whereas for the regular shifter circuit it is decreased by small margin due to the presence of extra XOR gate for switching the switchable FF. For simulation, the supply voltage (Vdd) for both the cases was considered 1.1 Volts

III. PIPELINE ADDERS

Pipelining is one of the technique which can be used to increase the throughput of a sequential set of distinct data inputs. For implementing pipelining requires including registers at each stage of the design. However, adding of the extra register for creating pipeline consumes more area and increases the latency. Different techniques were used to reduce the area and latency in the pipelined adder architecture. The overlapping clock is used on a conventional parallel pipelined adder [19] to reduce the sources of overhead [20]. For achieving high throughput, Lin *et al.* [21] uses proper scheduling a cascaded pipeline to eliminate data hazards. Carry propagation delay is one of the leading speed limiting factor in an adder design. For reducing the carry propagation delay across the pipelined adder, various techniques have being used [22], [23].

### A. Modified 4-bit pipelined parallel adder

To speed up the execution time, we implemented switchable DET register in a 4-bit pipelined parallel adder [19] and compared the results obtained in both the cases.

A 4-bit pipelined parallel adder is shown in Figure 3. As can be seen in the block diagram in Figure **3** the register pair was replaced with the switchable SET FF to DET FF.

Latency for M='1' is one clock period when the shifter circuit was shifting the capture data at both the clock edges. Whereas, for M='0' it is equal to 2 clock period which is same as the conventional 4-bit adder. This clearly present a decrease in the latency for M='1'. The latency is reduced by 50%. We implemented the design using 45nm PD-SOI standard cell library [21] to compare the area overhead, power consumption, and delay. Table 2 summarizes the numerical result for the 4-bit adder with the regular register and modified register for M='0' and M='1'.

### B. Simulation Result

The propagation delay presented in Table 2 is from clock to the output terminal of the switchable register pair. So the propagation delay or C2Q delay for M='0' and conventional pipeline adder are same. The propagation delay for M='1' is reduced by 95.1 % compared to conventional 4-bit parallel pipelined adder.

The power consumption was increased by 9% due to the addition of XOR gates for making register switchable. Whereas, the area for implementing 4-bit adder was increased by 1.365% due to additional XOR gates. The power delay product for M='1' was decreased by 44.16 % compared to the convention 4-bit adder.

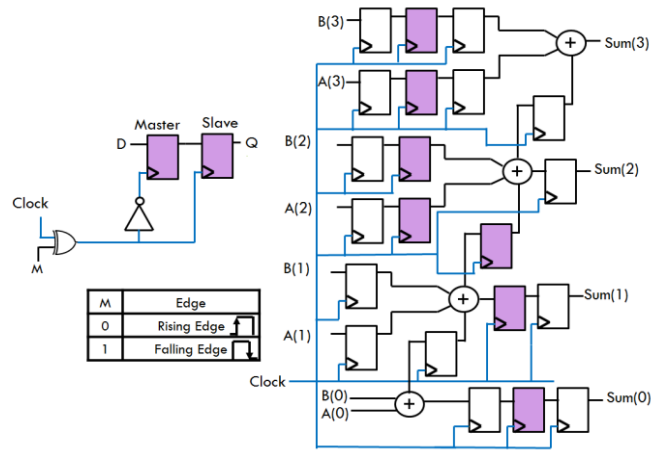

Figure 3 Modified 4-bit pipelined parallel adder

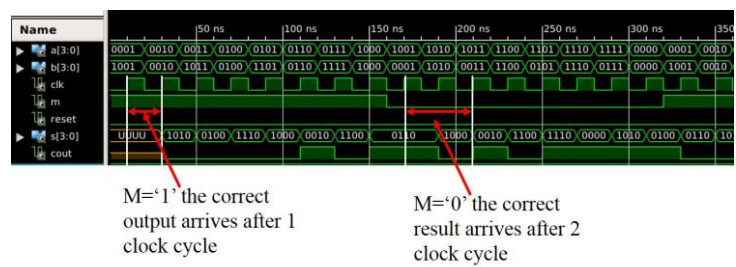

Figure 4 Circuit simulation result of modified 4-bit parallelism adder

Table 2 Simulation result

|  | Area ($\mu m^2$) | Delay (ns) | Power ($\mu W$) | Power Delay Product (fs-W) |
|---|---|---|---|---|
| **Pipeline Adder [19]** | 293 | 2.0069 | 24.26 | 48.7 |
| **M= '0'** | 297 | 2.0069 | 26.471 | 53.1 |
| **M= '1'** | 297 | 1.0286 | 26.471 | 27.2 |

IV. CONCLUSION

A switchable register is proposed in this paper as DET FF consumes more power than SET FF during high input activity. The switchable register can be made to switch from SET FF to DET based on load condition. The proposed register reduces the latency in 4-bit pipeline adder circuit by 50% compared with conventional pipeline adder. This technique also reduced the power delay product by 44.16%.